# First-light images from low dispersion spectrograph-cum-imager on 3.6-meter Devasthal Optical Telescope

**Amitesh Omar, T. S. Kumar, B. Krishna Reddy, Jayshreekar Pant, Manoj Mahto**
Aryabhatta Research Institute of Observational Sciences (ARIES), Manora Peak, Nainital, 263 001 India.
email: aomar@aries.res.in

A low dispersion spectrograph-cum-imager has been developed and assembled in ARIES, Nainital. The optical design of the spectrograph consists of a collimator and a focal reducer converting the f/9 beam from the 3.6-m Devasthal optical telescope to a nearly f/4.3 beam. The instrument is capable of carrying out broad-band imaging, narrow-band imaging and low-resolution ($\lambda/\Delta\lambda<2000$) slit spectroscopy in the wavelength range 350-1050 nm. A closed-cycle cryogenically cooled charge-coupled device camera, also assembled in ARIES, is used as the main imaging device for the spectrograph. The first images from the spectrograph on the telescope assert seeing-limited performance free from any significant optical aberration. An *i-band* image of the galaxy cluster Abell 370 made using the spectrograph shows faint sources down to ~25 mag. The quality and sensitivity of the optical spectrums of the celestial sources obtained from the spectrograph are as per the expectations from a 3.6-m telescope. Several new modes of observations such as polarimetry, fast-imaging, and monitoring of the atmospheric parameters are being included in the spectrograph. Using a test setup, single optical pulses from the Crab pulsar were detected from the telescope. The spectrograph is one of the main back-end instruments on the 3.6-m telescope for high sensitivity observations of celestial objects.

Introduction

A low dispersion spectrograph-cum-imager (*hereafter spectrograph*), following the design concepts of various Faint Object Spectrograph and Camera (FOSC) type instruments[1-4], has been developed for the 3.6-m Devasthal Optical Telescope (DOT). The spectrograph, designed and developed with mostly in-country research and development efforts, was assembled in ARIES, Nainital. The FOSC-type instruments are proven to be highly versatile and scientifically productive for low-medium dispersion spectroscopy and imaging of faint celestial sources. The spectrograph consists of a fixed collimator followed by a fixed focal-reducer unit with provisions for inserting optical filters and dispersive elements such as grism, prism, and gratings in the optical path. The filters, slits, and grisms etc. are mounted on different motorized rotating wheels for automated and fast operation via a remote computer. If slit and grism are not inserted in the optical path, the instrument works in non-dispersing imaging mode with focal reduction. In this way, the instrument's operation mode can be changed to imaging or spectroscopy within a few seconds by inserting or removing appropriate optical elements in the path. The instrument can also be made to work in the polarization mode if polarizing elements are inserted in the optical path.

The 3.6-m DOT, a moderate but the largest aperture optical telescope in India, is situated at Devasthal (Nainital) at the geographical location of $29.36^0$ N, $79.69^0$ E, and ~2426 meter above the mean sea level in the Himalayan regions of Uttarakhand. The DOT uses an f/9 Ritchey-Chrétien (RC) catoptrics system providing the Cassegrain image plane. The telescope points and tracks a celestial source using the altitude over azimuth mount. The telescope's primary mirror supported on 69 axial actuators uses the active optics technology, which was originally developed by European Southern Observatory in 1980s for the new technology telescope[5]. The technical details of the DOT and the Devasthal site are provided elsewhere[6-8]. The DOT has an internal auto-guider to improve tracking accuracy and a wave-front sensor to align the telescope optics. The 3.6-m DOT is suitable to carry out astronomical observations in many areas of Galactic and extra-galactic astronomy for observations of star clusters, young stars, supernova and gamma-ray bursts, variability of stars, galaxies, high redshift sources, and active galactic nuclei[9]. Most of these studies require imaging capabilities over several arc-minutes of Field of View (FoV) and spectroscopic capabilities over the visible waveband with spectral resolving power ($\lambda/\Delta\lambda$) in the range of 100–2000. This spectral resolving power is suitable for detecting the majority of spectral lines from faint celestial objects. The spectrographs with higher spectral resolving power, although limited to observations of relatively brighter objects, are also required to fulfill several other science aims envisaged for the DOT[9].

It was decided during the early phases of the DOT project that a low-dispersion optical spectrograph could be built using the in-country expertise. Subsequently, the design was started in ARIES with the helps from the experts in Satellite Application

Center (SAC) of the Indian Space Research Organization (ISRO) at Ahmedabad. After several rounds of design optimization in collaboration with the international experts and optical industries, final designs were made and manufacturing of the optics and the mechanical components were started. The main optics was fabricated, aligned, and tested in the premises of Winlight Systems, France. The precision mechanical components were fabricated and tested in the premises of Pawan Udyog and other small-scale industries in Ghaziabad-Delhi area as well as using the modern facilities at ARIES. The optics for the calibration unit and wedge-prisms were fabricated in the premises of Hindustan Opticals, Dehradun. The standard grisms and filters were imported from the reputed vendors such as Newport Richardson Gratings, U.S.A. and Asahi Spectra, Japan. The precision air-slits were manufactured using the laser processing facility at International Advanced Research Center for Powder Metallurgy and New Materials (ARCI), Hyderabad. The spectrograph was fully assembled and tested in ARIES. This is the first FOSC-type large-size astronomy instrument designed, developed and assembled in India. To record images from the spectrograph, a 4096 x 4096 pixel charge-coupled device (CCD) camera with an image area of 61.4 x 61.4 mm$^2$ was also designed, developed, and assembled in ARIES in technical collaboration with Herzberg Institute of Astrophysics (HIA), Canada. This paper presents a technical summary and the first-light images obtained from the spectrograph. Future up-gradation plans are also briefly presented. A comprehensive characterization report and technical details will be published elsewhere.

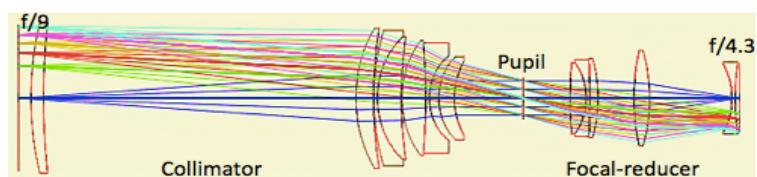

**Figure 1.** The ray-diagram of the spectrograph optics. The input is f/9 focal plane from the 3.6-m DOT and the output is ~f/4.3 focal plane imaged on the CCD camera. Total optical track length between the two focal-planes is ~1150 mm and the largest diameter of the lens used is ~200 mm.

**Main components and features of the spectrograph**

The spectrograph, a dioptric instrument, consists of a collimator, a focal-reducer and four motorized wheels accommodating the inter-changing optical elements, viz., slits, grisms/prism/grating, broad-band filters, and narrow-band filters. A spectral calibration unit is also part of the spectrograph. The optical design of the spectrograph is optimized to obtain 80% encircled energy within 0.4-arcsec diameter anywhere at the image plane within the central ~10-arcmin diameter. The length of the optical track from the telescope focal plane to the CCD sensor is ~1150 mm. A preliminary design concept of the spectrograph was presented elsewhere[10]. The ray diagram of the optics is shown in Figure 1. The final optical design uses several optical elements made of Calcium Fluoride (CaF2) and *i-line* high homogeneity glasses S-FSL5, PBM18Y, PBM2Y, PBL1Y, and BAL15Y. The *i-line* glasses have low birefringence and provide high internal transmission (~98% for 10 mm thickness) down to 365 nm wavelength. The spectrograph converts the f/9 optical beam from the 3.6-m DOT to a faster beam of ~f/4.3. The collimator unit consists of 7 lenses in five groups with 3 singlets and 2 doublets. The focal-reducer unit consists of 5 lenses in 3 groups with 1 triplet and 2 singlets. The last element, a field flattener made of fused silica, is mounted as the window of the CCD vacuum dewar. The field lens in the collimator unit is the largest optical component with a diameter of ~200 mm. The diameter of the pupil plane is ~45 mm. The optical surfaces are polished to a surface accuracy of ~5 nm rms to minimize light scattering. All the air-glass contact surfaces are also coated with a broad-band anti-reflective (AR) coating to minimize the light losses from the optical surfaces. The AR coating has reflectivity below 1.5% between 350 nm and 900 nm.

The broad-band color filters are mounted in the collimated beam before the pupil-plane providing a maximum usable FoV of 13.6 x 13.6 arcmin$^2$ on the 61.4 x 61.4 mm$^2$ CCD sensor. Presently, the Sloan Digital Sky Survey (SDSS) standard *ugriz* filter system is mounted in five slots. The classical *UBVRI* filter system will also be made available in near future. There are a total of ten slots in the broad-band filter wheel, which can accommodate five 80 mm and five 90 mm diameter filters. The grisms are mounted at the pupil plane. The grisms of 50 mm x 50 mm size have transmission gratings with 300–600 lines/mm. The spectral dispersion is in the range 0.1–0.23 nm per 15 micron pixel (~0.2 arcsec in sky) or the effective spectral resolving power at 1 arcsec of atmospheric seeing is nearly in the range 500–1000. At the atmospheric seeing near 0.5 arcsec, a resolving power of 2000 may be obtained. An 11-degree wedge prism is also available to get very low dispersion (~6 nm per 15 micron) spectrum suitable for slit-less spectroscopy. We are in process of procuring additional gratings at lower resolutions and virtual phase holographic grisms to further improve the sensitivities. The narrow-band filters are mounted in the f/9 beam before the telescope focal-plane providing a clear FoV of ~11 arcmin diameter. The narrow-band filter wheel can accommodate five 120-mm filters. This wheel also has a clear slot of ~192 mm diameter to enable the light-cone for 13.6x13.6 arcmin$^2$ FoV to pass through it un-obstructed for broad-band imaging. The slits are mounted at the f/9 focal plane of the telescope. The slits provide ~8 arcmin long FoV with widths ranging between 0.4 and 2 arcsec. The slits were cut on 100 μm thick SS-304 and 20 μm thick brass foils using the precision laser cutting machine at ARCI, Hyderabad. The electronics control hardware for rotating each of the four wheels via anti-backlash drive mechanism was designed and developed at ARIES with capabilities for remote operation. The wheels and the calibration lamps are operated and controlled using a graphical user interface.

The spectrograph works in imaging and spectroscopy modes. When the photometric imaging observations are desired, only filters are inserted in the optical path while when the spectroscopic observations are desired, grisms and slits are inserted in the optical path. As the bandwidth (~700 nm) of the usable spectrum (350-1050 nm) covered by the spectrograph is more than the lowest wavelength (~350 nm) of interest, second order spectrum in the blue region (350-700 nm) gets superimposed on the first order spectrum in the red region (700-1050 nm). The lower cutoff at ~350 nm is due to Earth's atmosphere, which does not pass significant light below it and the upper cutoff at near 1050 nm is due to sensitivity cutoff of the silicon based CCD sensor. Two filters BG-39 (passing 350-700 nm) and RG-610 (passing >600 nm) are used as the spectral-order blocking filters to block second or first order spectrum from the grating. Either of these filters can be used to obtain pure spectral order dispersion either in the blue or the red region without any contamination from the other order.

An in-built calibration unit enables spectral calibration of the dispersion axis on the CCD detector and also to carry out flat-fielding of the spectrum. The Hg-Ar and Ne lamps are used for spectral calibration while a Tungsten-Halogen lamp emitting continuum light is used for the flat-fielding. A light-integrating sphere is used to get a near uniform light source from the light coming from the lamps. The calibration unit has a biconvex lens, which makes the light from the integrating sphere to converge with a focal ratio of ~f/9, same as that of the telescope. The converging light path from the lens is perpendicular to the telescope light path and hence a 45-degree mirror is used to deflect the converging light from the lamps to focus on the slit-plane. The lens and the 45-deg mirror of the calibration unit are mounted on two separate motorized linear stages to adjust the focus. When a calibration observation is desired, the lamps are switched on and the lens-mirror system is positioned to focus light on the slit plane and no light from the sky is received as the 45-degree mirror blocks the sky. When science observations are desired, the lens-mirror system is retracted away from the telescope's light path and the lamps are switched off.

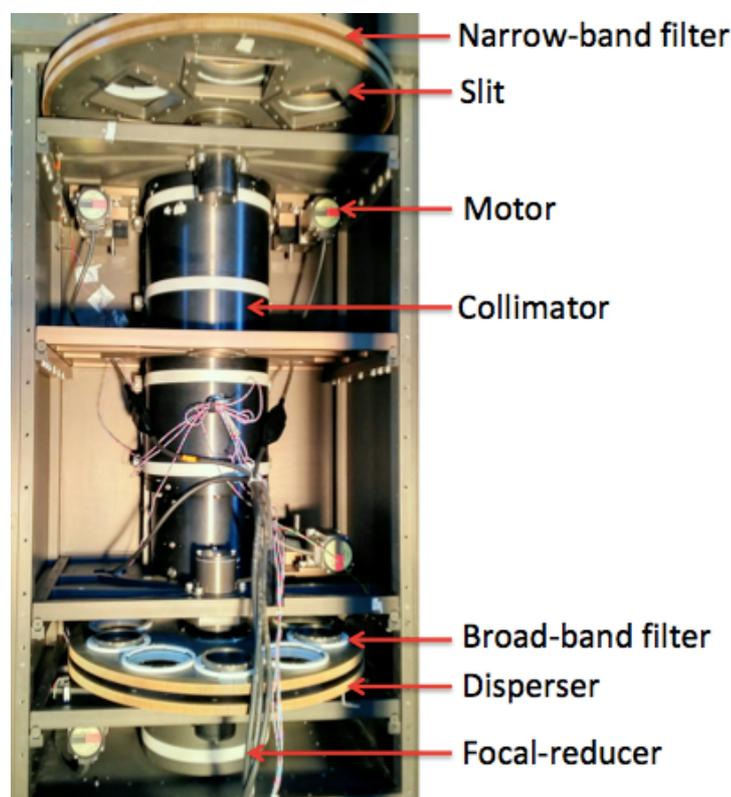

**Figure 2.** An internal view of the assembled spectrograph showing major components. The light from the telescope enters from the narrow band filter side and image is formed behind the focal reducer.

The optics and other components are mounted inside a precision-machined enclosure made from the Aluminum-alloys 6061 and 6063. The main optics (collimator and focal-reducer) is mounted on a V-grooved block. The enclosure is black anodized to minimize light scattering. The weight of the spectrograph is nearly 500 kg. Since the spectrograph is mounted directly on the telescope at one end with another end free to bend due to gravity, provisions are made to arrest mechanical flexure by providing strengths to the structure using beams and heavy-duty turn-buckle fixtures. Since the balancing of the telescope's altitude-axis tube requires a fixed torque, an additional dummy weight of nearly 1300 kg is also attached along with the spectrograph. This additional weight, attached to the telescope-mounting flange, is structurally detached from the spectrograph and has provisions for adding or removing weight plates to fulfill the balancing requirements. In order to mount and dismount the spectrograph and the dummy weight structure on the DOT, a special integration trolley was designed at ARIES and fabricated at small-scale industries in Ghaziabad-Delhi area. The internal view of the instrument's assembly and a picture of the instrument as mounted on the 3.6-m DOT are shown in Figures 2 and 3 respectively.

A CCD camera was also designed and assembled in ARIES for the spectrograph. The CCD camera was developed in collaboration with HIA, Canada. The camera uses a 4096x4096 format grade-0 back-illuminated E2V 231-84 CCD sensor having square pixels of size 15 micron. The CCD sensor is a deep-depletion Silicon device with multi-layer broadband coating optimized for fringe-free imaging in the red side. The quantum efficiencies of the CCD are ~84% at 400 nm and ~60% at 900 nm with peak value of ~90% near 650 nm. The CCD has quad-amplifier architecture and its ~16 million pixels can be read from any of the four amplifiers or through four amplifiers each reading ~4 million pixels from the four quadrants. The read-out frequency is fixed at ~160 kHz to keep the read-out electronic noise below 10 e- rms using a controller from Astronomical Research Cameras, Inc. USA. The CCD sensor is cooled to -120$^0$C using a closed-cycle (Joule-Thompson) cryogenic heat-exchange system supplied by Brooks Automation, USA. The temperature of the sensor is stabilized and held constant within 0.01$^0$C using a proportional-integral-derivative controller and a small heater below the sensor mounting plate. The CCD is operated in non-inverted mode. With these settings, the per-pixel dark noise generated is extremely low at a value ~3 e- per hour. The sensor and a circuit-protection board are mounted inside a vacuum dewar. A charcoal filled getter is used to absorb outgassing inside the dewar. The getter gets activated at cryogenic temperatures and helps in attaining high vacuum inside the dewar. The dewar attains a vacuum ~3x10$^{-7}$ Torr when its internal system and the CCD sensor are cooled to -120 $^0$C. The closed-cycle cooling system with heat-exchanger pipes, although somewhat cumbersome to use on an altitude-azimuth telescope mount, alleviate the need to regularly fill the liquid nitrogen commonly used in the CCD cameras for astronomy. The closed-cycle cooling system also makes the CCD dewar compact and less bulky as the liquid nitrogen holding tank is not needed in such systems. Further details of the CCD system and its characterization will be published elsewhere.

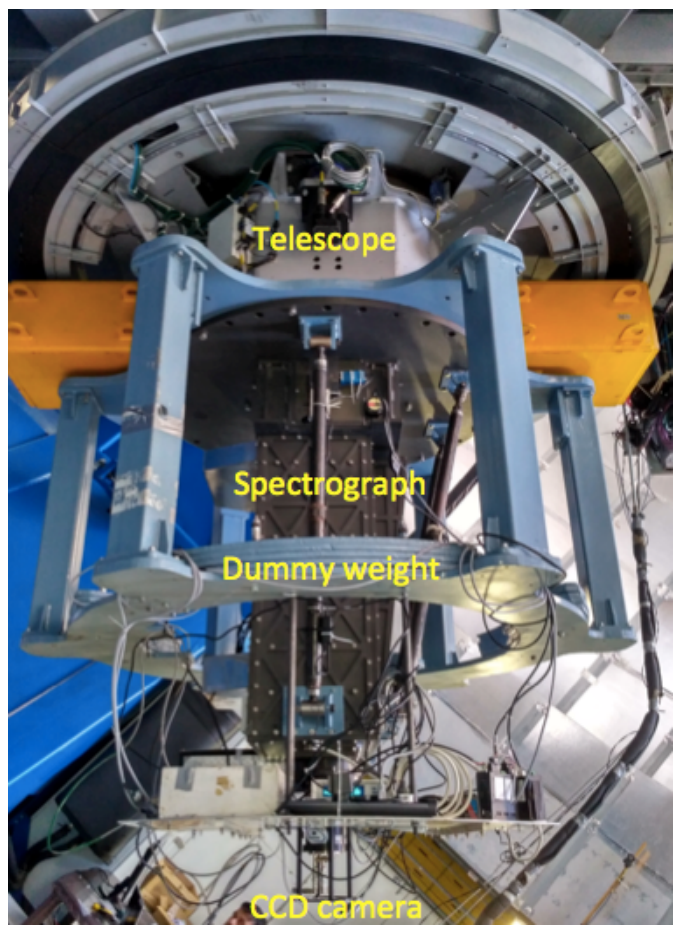

**Figure 3.** A picture of the spectrograph mounted on the 3.6-m DOT.

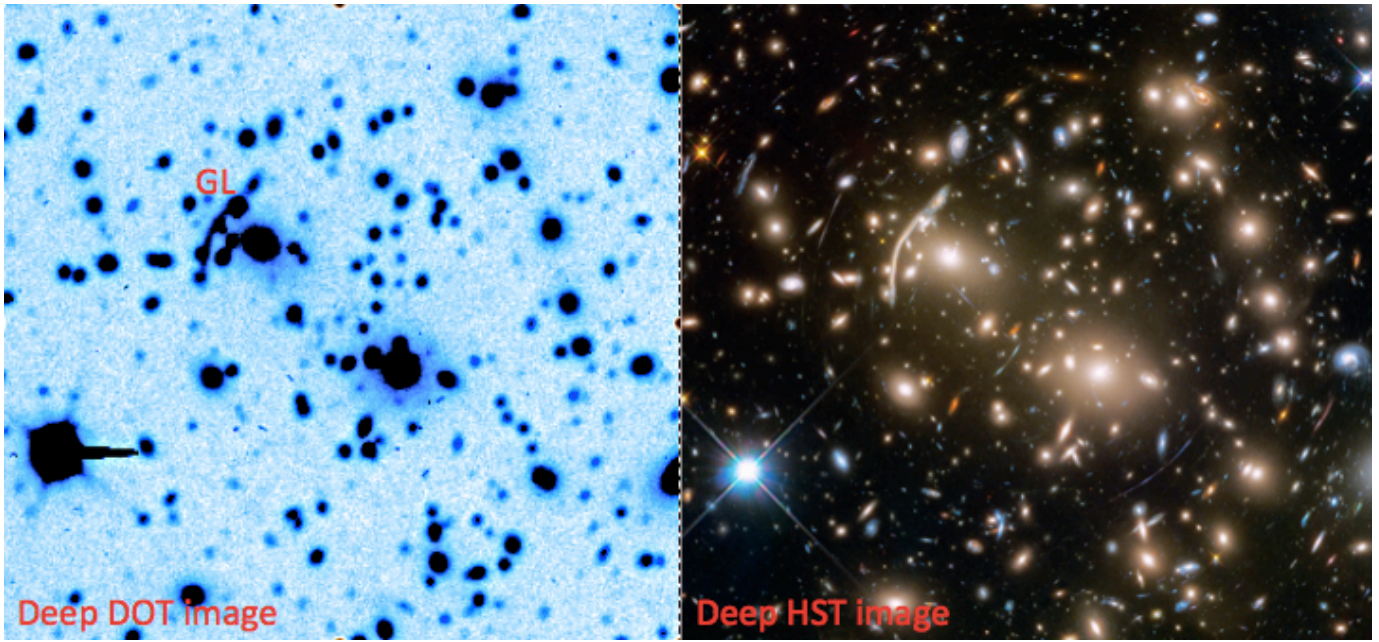

**Figure 4.** (left) An *i-band* image of the central region (~3'x3') of the galaxy cluster Abell 370 made using the observations from the spectrograph on the 3.6-m DOT in the imaging mode. (right) A deep image made using the HST (credit: NASA, ESA, J. Lotz and the Hubble Frontier Fields team of STScI) is reproduced to make a comparison of the locations of the faint detected sources. Most of the diffuse objects in the image are galaxies. The detection limit in the DOT image of 55 min exposure is at ~25 mag with a photometric precision of 0.3 mag. The prominent gravitational-lens arc is marked as GL.

**On-sky performance:**

The spectrograph has been mounted on the 3.6-m DOT a few times to analyze the on-sky performance. The instrument was used in both imaging and spectroscopy modes for the observations of stars, ionized star-forming regions and galaxies. Figure 4 presents a calibrated image of the galaxy cluster Abell 370 in the SDSS *i-band* (700- 850 nm) with total exposure time of 55-minute taken in 11 frames of 5-minute each on 14-Nov-2017. The photon counts in the image were converted to flux (magnitude) using the calibrated image of the same region available from the SDSS archive. The image has FWHM of ~1.5 arcsec and the image frames were taken when the object was in the West near the zenith distance of ~$55^0$. The *i-band* sky brightness at Devasthal at the time of the observations determined from this image is ~20 mag arcsec$^2$, which is at par with several dark observing sites worldwide. The background-sky is uniform (flat) in the processed image to a high accuracy of ~0.1% rms. The faintest sources detected in this image are i~24-25 mag with signal-to-noise ratios between 3 and 5. The locations of the faintest sources in this image match with those in the deep image of the same region, available from the Hubble Space Telescope (HST) legacy program. The famous giant gravitational lens arc is also clearly visible in this image. A 25-mag source has a feeble effective photon rate of about 1 photon/sec through the spectrograph on the 3.6-m DOT. To put this faint detection in perspective, human eyes can detect a star of about 7 mag, which has about 10 million times higher photon rate than a 25 mag source detected here. The best-recorded stellar images from the spectrograph have FWHM ~1.5 arcsec in the *u*-band (~350 nm) and ~1.1 arcsec in the *z*-band (~900 nm). The high fidelity in the DOT images and excellent correspondence in terms of faint detections with the HST images imply that the spectrograph on the 3.6-m DOT is capable of making very deep images of faint sources to a high accuracy. It is worth to mention here that as the CCD camera used in this spectrograph is sensitive in the red side, scientific observations of the near-infrared bright sources such as high-redshift galaxies and extremely cold or low-mass stars (white dwarfs, brown dwarfs etc.) using this spectrograph will also be possible.

Figure 5 shows the calibrated spectrum of the extra-galactic supernova (SN) 2017gmr made using the observations from the spectrograph taken on 17-Nov-2017, when the supernova was only about 73 days old. At the time of the observations, the supernova has the visual magnitude of ~15. The line-strengths of the prominent emission-lines of the elements H, C, N, O, Fe as marked in Figure 5 indicate that this supernova is Hydrogen rich (Type-II) and most likely produced from the explosion of a runaway massive star. It is important to mention that this spectrum was created in a single exposure of 5-minute duration, covering the entire wavelength range between 400 nm and 900 nm. This spectrum has an excellent correspondence in terms of the line features and strengths with the archived spectra of similar type of supernovae. The supernovae at this stage evolve rapidly and the line features change every few days. The spectroscopic observations of supernovae over regular intervals provide crucial information on evolution and nucleo-synthesis of various elements in the Universe. The high fidelity of the optical spectrum indicates that this spectrograph on the 3.6-m DOT will be very valuable for spectroscopic studies of faint celestial objects.

**Possible up-gradations:**

Several new modes of observations are possible to include in the FOSC-type spectrograph presented here. A polarimetry mode is planned to be included using a Wollaston prism and an appropriate wave-plate. This mode will enable measurement of polarization of the light coming from various celestial objects. It is also possible to include multi-object spectroscopy mode using a slit mask customized for a field. A fast-imaging mode to measure small variations in the flux of variable sources at millisecond cadence is being included in the present design. For this mode, a frame-transfer electron-multiplying (EM) CCD camera is presently being tested on the 3.6-m DOT. This CCD camera provides a FoV of 1.7 x 1.7 arcmin$^2$ with readout rate up to ~350 frames-per-second. A Global Position System (GPS) based time event monitoring system expected to make time-stamping of the image frames at an accuracy <0.1 ms is also presently under testing. A test observation made using the EM-CCD camera directly on the 3.6-m DOT fetched exciting results on the observations of the Crab pulsar (PSR B0531+21), which is a young neutron star spinning rapidly with a rotation period of ~33.7 ms. Figure 6 shows repeating optical pulses from the Crab pulsar, detected from the 3.6-m DOT. Each data point here corresponds to ~3.37 ms of effective cadence (exposure + read-out time). A sensitivity estimate indicates that a ~17 mag source can be expected to get detected in 10 ms integration using a high-sensitivity EM-CCD camera. This set-up can also be effectively combined with a low dispersion prism and an appropriate multi-band-pass filter to carry out multi-color imaging of some bright (<17 mag) sources at fast cadence similar to a multi-channel fast photometer.

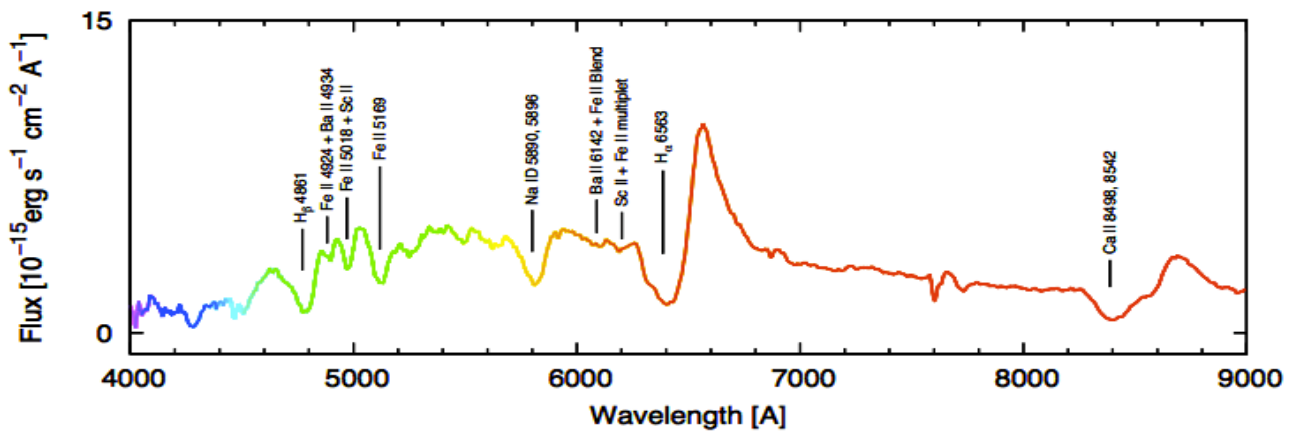

**Figure 5.** The optical spectrum of the supernova 2017gmr made using the spectrograph on 3.6-m DOT in a single exposure of 5 min. The prominent elemental lines are marked. The spectrum has average resolution (Δλ) of ~10 Angstrom. The rms noise values (in erg/cm$^2$/s/ Å) are 4x10$^{-16}$ near 4500 Å, 2x10$^{-16}$ near 6000 Å, 0.8x10$^{-16}$ near 7500 Å and 1x10$^{-16}$ near 9000 Å. The peak detections of the Hα and Hβ lines are made at signal-to-noise ratios of 70 and 20 respectively.

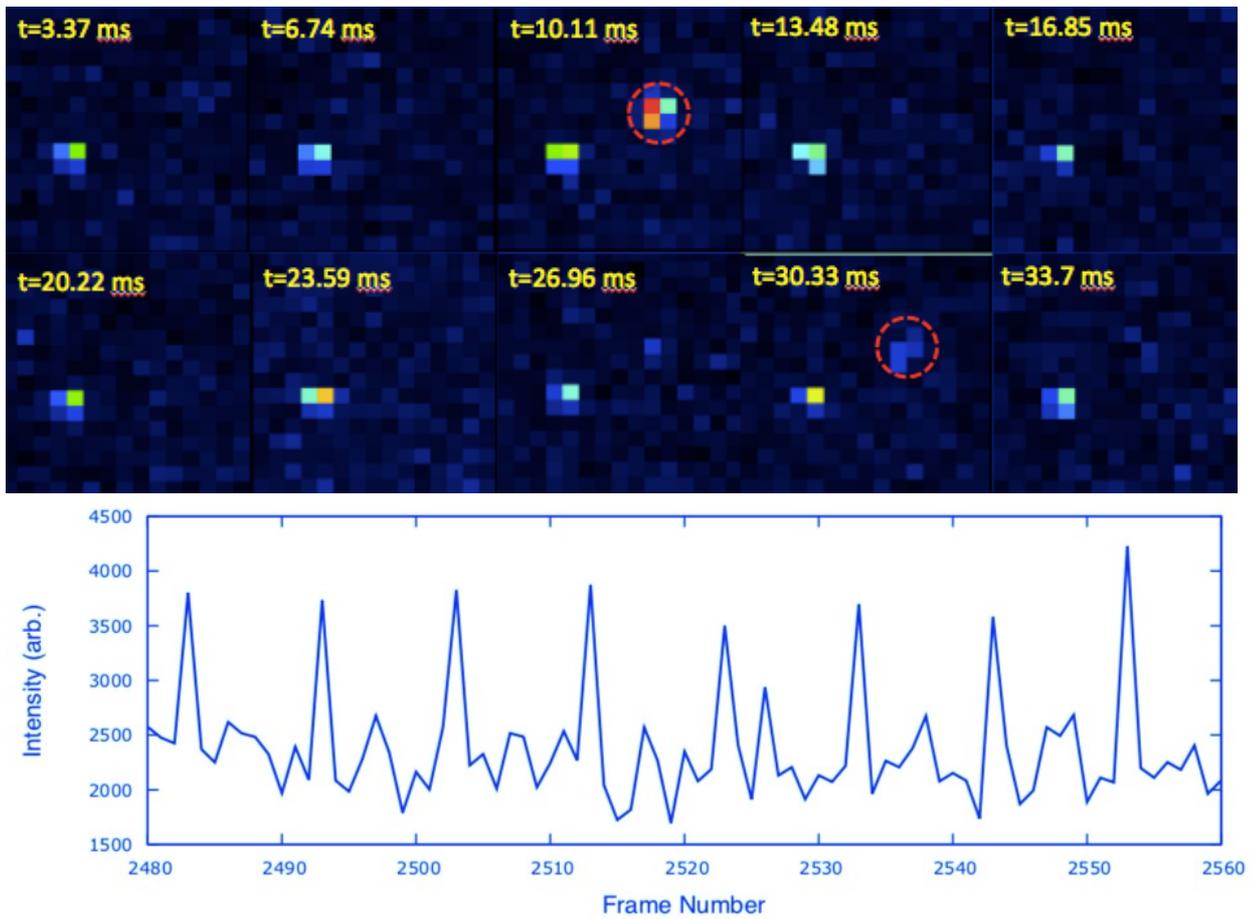

**Figure 6.** (top) The milli-second images for one cycle of pulse emission from the Crab pulsar, marked by red circles, recorded using an EM-CCD camera mounted on the 3.6-m DOT. The object with persistent emission to the left of the pulsar is a star. (bottom) The intensity profile of the optical pulses from the Crab pulsar. The abscissa represents frame numbers with time separation of ~3.37 ms. The pulsar has a known period of ~33.7 ms. Both strong main pulse (t~10 ms) and weak inter-pulse (t~27 - 30 ms) are detected.

The annular FoV beyond the central 13.6 x13.6 arcmin$^2$ used by the spectrograph is also available at the f/9 beam as the 3.6-m DOT has a total usable FoV of ~30 arcmin diameter at the Cassegrain focal plane. This annular FoV can be imaged by another CCD camera directly in the f/9 beam to provide an independent measurement of the sky parameters such as extinction. This CCD camera can also be used for external guiding to improve tracking accuracy. A set-up using a 0.5-degree wedge prism at the pupil plane is also being tested in the spectrograph to measure atmospheric seeing parameters based on Differential Image Motion Monitor (DIMM) principles[11]. This DIMM set-up and continuous extinction measurement set-up within the spectrograph will be valuable for getting accurate photometry of the celestial objects.

**Conclusions:**

A spectrograph similar to the popular and versatile astronomy instrument FOSC has been designed and developed in ARIES. This is the first large-size FOSC-type astronomy spectrograph designed and assembled within the country, thanks to the expertise available in various Indian government institutes and organizations, industries, and technical support received from the organizations abroad most notably from Belgium, Canada and France. A large format closed-cycle cryogenically cooled CCD camera was also developed and assembled in ARIES for the spectrograph. The spectrograph and the CCD camera were mounted on the 3.6-m DOT for conducting various tests. The first-light results are showing satisfactory performance of the spectrograph. The images as deep as ~25 mag in the SDSS *i-band* with excellent uniformity across the full FoV have been made. The spectroscopy was successfully performed on faint celestial objects. The individual optical pulses from the Crab pulsar were detected using a fast EM-CCD camera on the telescope. In future, the fast read-out mode coupled with a GPS timing system in the spectrograph will enable precise multi-wavelength coordinated optical follow-up observations of fast transients along with other telescopes such as upgraded Giant Meterwave Radio Telescope (GMRT)[12] in radio wavebands and AstroSat[13] in high-energy bands. A polarimetry mode and other features such as seeing and extinction monitor and external guiding are also planned to be included in the spectrograph to further enhance its capabilities.

**Acknowledgements:**

The assembly of this instrument could not have been possible without the helps received from several organizations, industries, and individual experts. We acknowledge academic, technical, and administrative supports received from the staff of ARIES, too many to name here individually, to build this instrument and get first light on it through the 3.6-m DOT. The supports received from SAC-ISRO, Ahmedabad and ARCI, Hyderabad are duly acknowledged. The supports received from the Department of Science and Technology (DST), Government of India, ARIES Governing Council and Project Management Board of the 3.6m DOT project are duly acknowledged. We thank the members of the review committees of the project to provide constructive inputs at various stages. We thank A. Paswan, R. Dastidar and Dr. K. Misra for their help during the observations and in data reduction. We thank the DOT operation team led by Dr. Brijesh Kumar and director Dr. Wahab Uddin for their inputs and supports. We thank Profs. P.C. Agrawal, R. Sagar, S. N. Tandon, G. Srinivasan, S. Ananthakrishnan, T. P. Prabhu, G. Swarup, S. K. Joshi, and P. Asthana for the encouragements to develop the spectrograph within the country. We acknowledge discussions and supports from R.K.S. Yadav, S. Mondal, V. Shukla, S. Yadava, N. Nanjappa, T. Bangia, A. Misra, A.N. Ramaprakash, G. Raskin, J. Surdej. T. Hardy, and G. Burley. We are thankful for the timely support received from HIA, National Research Council, Canada to develop the CCD camera. We are thankful to Profs. G. Padmanabhan and R. Bathe for the timely support to develop slits in ARCI, Hyderabad.

**References:**

1. Buzzoni, B. *et al.*, The ESO Faint Object Spectrograph and Camera (EFOSC). *ESO Messenger*, 1984, **38**, 9-13.
2. Andersen, J. *et al.*, New power for the Danish 1.54-m telescope. *ESO Messenger*, 1995, **79**, 12-14.
3. Kashikawa, N. et al., FOCAS: The Faint Object Camera and Spectrograph for the Subaru Telescope. *Publications of the Astronomical Society of Japan*, 2002, **54**(6), 819-832.
4. Hook, I. M. *et al.*, The Gemini-North Multi-Object Spectrograph: Performance in Imaging, Long-Slit, and Multi-Object Spectroscopic Modes. *The Publications of the Astronomical Society of the Pacific*, 2004, **116** (819), 425-440.
5. Hardy, John W., Active optics: A new technology for the control of light. *Proc. of the IEEE*, 1978, **66**, 651.
6. Ninane, N., Carlo, F. and Kumar, B., The 3.6 m Indo-Belgium Devasthal optical telescope: general description. *In Proc. of the SPIE*, 2012, **8444**.
7. Kumar, B. *et al.*, 3.6-m Devasthal Optical Telescope Project: Completion and first results. *Bulletin de la Société Royale des Sciences de Liège (Actes de colloques),* 2018, **87**, 29 - 41
8. Sagar R. *et al.*, Evaluation of Devasthal site for optical astronomical observations. *Astronomy & Astrophysics Suppl.*, 2000, **144**, 349-362
9. Omar, A., Kumar, B., Gopinathan, M. and Sagar, R, Scientific capabilities and advantages of the 3.6 meter optical telescope at Devasthal, Uttarakhand. *Current Science*, 2017, **113**, 682-685.
10. Omar A., Yadav R. K. S., Shukla V., Mondal S. and Pant J., Design of FOSC for 360-cm Devasthal optical telescope. *In Proc. of the SPIE*, 2012, **8446**.
11. Sarazin, M. and Roddier, F., The ESO differential image motion monitor. *Astronomy & Astrophysics*, 1990, **227**, 294.
12. Gupta, Y. *et al.*, The upgraded GMRT: opening new windows on the radio Universe. *Current Science*, 2017, **113**, 707-714.
13. Singh, K.P. and Bhattacharya, D., Multi-color hues of the Universe observed with AstroSat. *Current Science*, 2017, **113**, 602-60